\documentclass{elsart}
\usepackage{graphicx}
\usepackage{amssymb}
\begin{document}
\begin{frontmatter}
\title{Magnetization plateaux in a generalised ladder model}
\author{Emily Chattopadhyay and Indrani Bose}
\address{Physics Department, Bose Institute, 93/1, A.P.C.
Road, Calcutta-700009, India}
\begin{abstract}
A spin-1/2 antiferromagnetic(AFM) generalised ladder model is constructed
which consists of four-spin plaquettes, coupled through weaker exchange
interactions, to two-spin rungs. In an extended parameter regime,
the exact ground state of the ladder is determined. In this state,
the four-spin plaquettes and the rungs are in their ground state spin
configurations. In the presence of an external magnetic field, the
magnetization/site has a plateau structure as a function of the magnetic
field.
\end{abstract}
\end{frontmatter}

\section{Introduction}

Recently, low-dimensional spin systems, particularly the Heisenberg
spin ladders, have been the focus of several analytic, numerical and
experimental studies{[}1,2{]}. Spin ladders can be considered as bridges
between one-dimensional (1D) and two-dimensional(2D) systems. The
1D systems are more or less well understood whereas considerable gaps
still exist in our understanding of 2D systems. The study of ladder
models is expected to provide insight on how electronic and magnetic
properties are modified from 1D to 2D. A large number of magnetic
compounds with ladder structure have been discovered{[}1,2{]} which
exhibit a variety of novel phenomena in the undoped as well as the
doped states. Different types of spin-1/2 ladder models have been
proposed including frustrated ladder models and models with modulated
exchange interactions{[}3-14{]}. Ladder models have been studied in
zero as well as finite magnetic fields. In this paper, we construct
a spin-1/2 ladder model with modulated exchange interactions. The
model consists of four-spin plaquettes connected to two-spin rungs.
The dominant exchange interactions are within the plaquettes and the
rungs. The coupling between the plaquettes and the rungs are through
weaker exchange interactions. Molecular magnets provide another example
of spin systems consisting of weakly coupled spin clusters{[}15{]}.
In the case of our ladder model, the spin clusters are the four-spin
plaquettes and the two spin rungs.

We show that in a certain parameter regime, the exact ground and low-lying
excited states of the full ladder model are of the product form, i.e.,
can be written in terms of the exact ground states of the four-spin
plaquettes and the two-spin rungs. Also, in the presence of an external
magnetic field, the magnetization/site exhibits the phenomenon of
magnetization plateaux. The condition for the appearance of a plateau
is given by{[}16{]}\begin{equation}
\label{1}
S_{u}-m_{u}=integer
\end{equation}
where \( S_{u} \) and \( m_{u} \) are the total spin and magnetization
in unit period of the ground state.


\begin{figure}[t]
\begin{center}
\includegraphics[width=6cm]{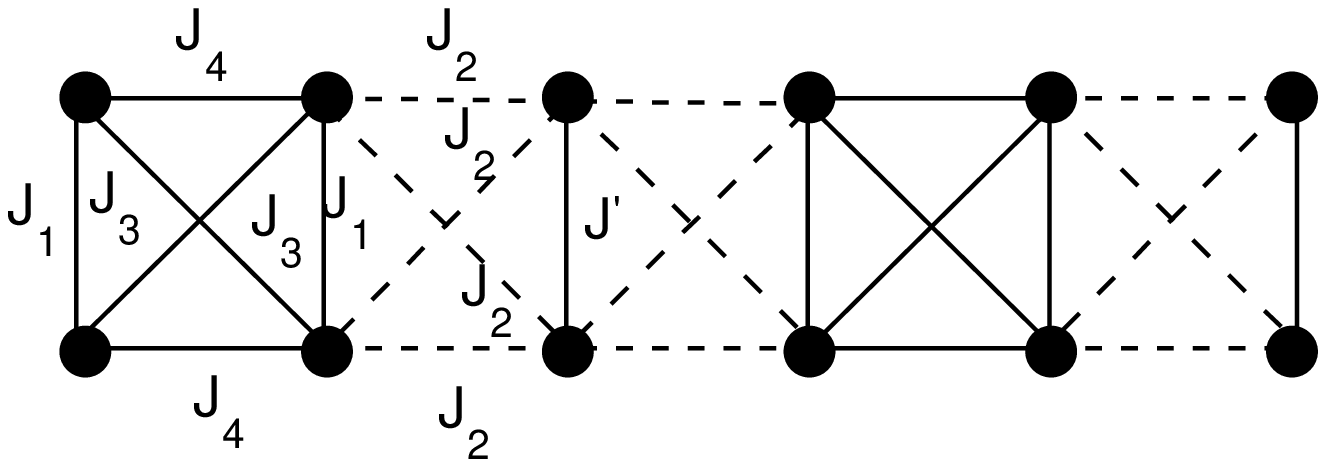}
\end{center}
\caption{Two-chain ladder model consisting of four-spin plaquettes
coupled to two-spin rungs. The exchange interaction strengths are
as shown in the Figure.}
\label{fig1}
\end{figure}


\section{Ground and excited states of the ladder model}

The ladder model constructed by us is shown in Fig. 1. The solid lines
represent the dominant exchange interactions within the four-spin
plaquettes and along the two-spin rungs. The dotted lines describe
the weaker exchange couplings between the plaquettes and the rungs.
Within a plaquette, the horizontal, vertical and diagonal exchange
interactions are of strengths \( J_{4} \), \( J_{1} \) and \( J_{3} \)
respectively. The exchange interaction along a two-spin rung is of
strength \( J^{\prime } \). The ladder model consists of alternating
four-spin plaquettes and two-spin rungs coupled via horizontal and
diagonal exchange interactions (dotted lines) of strength \( J_{2} \).
Periodic boundary condition is assumed to hold true. The ladder model
generalises a simpler model studied earlier{[}17{]} in which the horizontal
and vertical exchange interactions in a four-spin plaquette are of
equal strength. The spin Hamiltonian describing the ladder model is\[
H=\sum _{i=3j+1,j=0,1,\cdots }[J_{4}(\overrightarrow{S}_{1i}.\overrightarrow{S}_{1i+1}+\overrightarrow{S}_{2i}.\overrightarrow{S}_{2i+1})+J_{1}(\overrightarrow{S}_{1i}.\overrightarrow{S}_{2i}+\overrightarrow{S}_{1i+1}.\overrightarrow{S}_{2i+1})\]

\[
+J_{3}(\overrightarrow{S}_{1i}.\overrightarrow{S}_{2i+1}+\overrightarrow{S}_{2i}.\overrightarrow{S}_{1i+1})]+J^{\prime }\sum _{i=3j,j=0,1,\cdots }\overrightarrow{S}_{1i}.\overrightarrow{S}_{2i}\]

\begin{equation}
\label{2}
+J_{2}\sum _{i=3j+2,j=0,1,\cdots }(\overrightarrow{S}_{1i}+\overrightarrow{S}_{2i}+\overrightarrow{S}_{1i+2}+\overrightarrow{S}_{2i+2}).(\overrightarrow{S}_{1i+1}+\overrightarrow{S}_{2i+1})
\end{equation}

\[
=H_{C}+H_{R}+H_{CR}\]
The spin operator \( \overrightarrow{S}_{1i} \) (\( \overrightarrow{S}_{2i} \)
) is associated with the \( i \)-th site of the lower (upper) chain
of the ladder, the site indices are sequential in a chain. The sub-Hamiltonians
\( H_{C} \) and \( H_{R} \) describe the four-spin plaquettes and
the rungs, respectively, whereas \( H_{CR} \) contains the exchange
couplings between the plaquettes and the rungs. The total spin of
each rung is a conserved quantity due to the special structure of
the Hamiltonian.

We now determine the ground state of the ladder model using the method
of `divide and conquer'{[}18{]}. It is easy to show that the state,
in which the four-spin plaquettes and the two-spin rungs are in their
ground state spin configurations, is an exact eigenstate of the full
Hamiltonian H(Eq. 2). H is a sum of three sub-Hamiltonians \( H_{C} \),
\( H_{R} \) and \( H_{CR} \). The Hamiltonian \( H_{C}+H_{R} \)
acting on the specified state gives back the same state with the eigenvalue,
\( E_{CR} \), equal to the sum of the ground state energies of all
the plaquettes and the rungs in the ladder. The sub-Hamiltonian \( H_{CR} \)
acting on the same state gives zero. Thus, the state is an exact eigenstate
of H with the eigenvalue \( E_{1}=E_{CR} \). In an extended parameter
regime, the exact eigenstate also turns out to be the exact ground
state. The proof is as follows:-

Let \( E_{g} \) be the exact ground state energy of the full Hamiltonian
H. Then \( E_{g}\leq E_{1} \). Let \( \left| \psi _{g}\right\rangle  \)
be the exact ground state wave function. Then from variational theory,

\begin{equation}
\label{3}
E_{g}=\sum _{j}\left\langle \psi _{g}\right| H_{j}\left| \psi _{g}\right\rangle +\sum _{j}\left\langle \psi _{g}\right| H^{\prime }_{j}\left| \psi _{g}\right\rangle \geq \sum _{j}(E_{jo}+E^{\prime }_{jo})
\end{equation}

\[
H=\sum _{j}(H_{j}+H^{\prime }_{j})\]
where \( H_{j} \) 's are the plaquette Hamiltonians with the ground
state energy \( E_{jo} \) and \( H^{\prime }_{j} \)'s are the six-spin
cluster Hamiltonians, each of which contains the rung exchange interaction
Hamiltonian and the eight exchange couplings (four horizontal and
four diagonal) which connect the rung to nearest-neighbour plaquettes.
The ground state energy of \( H^{\prime }_{j} \) is \( E^{\prime }_{jo} \).
For \( J_{2}\leq \frac{J^{\prime }}{4} \), \( E^{\prime }_{jo} \)
is the ground state energy of the rung Hamiltonian. In the ground
state, the rung is in a singlet configuration. We can now write down
the inequality,

\begin{equation}
\label{4}
\sum _{i}(E_{io}+E^{\prime }_{io})\leq E_{g}\leq E_{1}
\end{equation}
 \( E_{1} \) is, however, exactly equal to \( \sum _{i}(E_{io}+E^{\prime }_{io})=E_{CR} \).
Thus, \( E_{g}=E_{1} \), i.e., the exact eigenstate of the ladder
model is also the exact ground state. The exact ground state energy
is given by \( E_{g}=N(E_{i0}-3\frac{J^{\prime }}{4}) \) where N
is the total number of plaquettes as well as rungs in the ladder. 


{\centering \textbf{TABLE I}\par}

\vspace{0.3cm}

\begin{tabular}{|c|c|c|cc|}
\hline 
S&
 Eigenvalues&
 S\( ^{z} \)&
&
 Eigenstates\\
\hline
\multicolumn{1}{|c|}{0}&
 \( -\frac{J_{1}+J_{4}+J_{3}}{2}-X \)&
 0&
\( \left| \psi _{1}\right\rangle  \)=&
 \resizebox*{1.5in}{.4in}{\resizebox*{1in}{0.5in}{\includegraphics{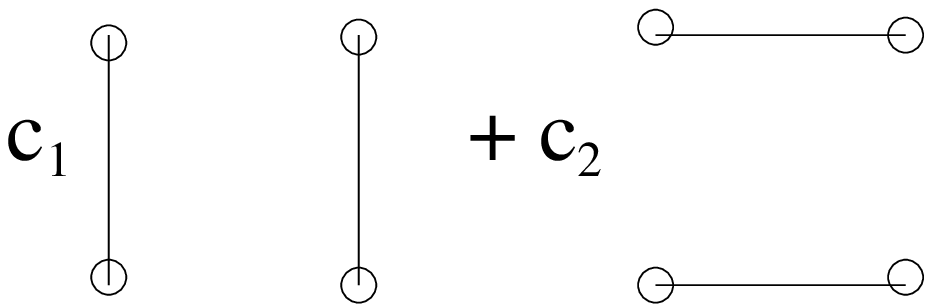}} }\\
\hline
0&
 \( -\frac{J_{1}+J_{4}+J_{3}}{2}+X \)&
 0&
\( \left| \psi _{2}\right\rangle  \)=&
 \resizebox*{1.5in}{.4in}{\resizebox*{1in}{0.5in}{\includegraphics{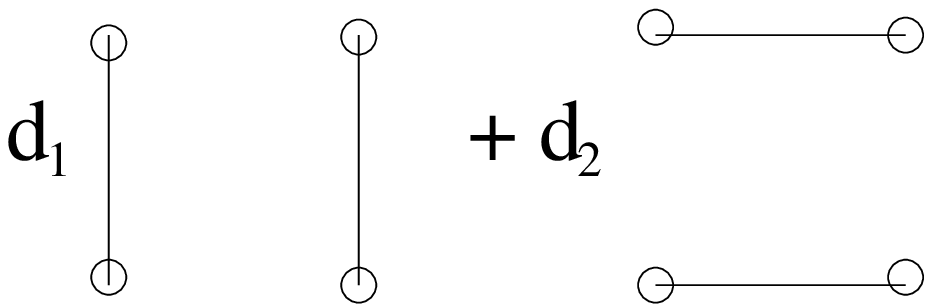}} }\\
\hline 
1&
 \( \frac{-J_{1}-J_{4}+J_{3}}{2} \)&
 0&
\( \left| \psi _{3}\right\rangle  \)=&
 \( \uparrow \downarrow \downarrow \uparrow -\downarrow \uparrow \uparrow \downarrow  \)\\
\hline
1&
 \( \frac{-J_{1}+J_{4}-J_{3}}{2} \)&
 0&
\( \left| \psi _{4}\right\rangle  \)=&
 \( \uparrow \downarrow \uparrow \downarrow -\downarrow \uparrow \downarrow \uparrow  \)\\
\hline
1&
 \( \frac{J_{1}-J_{4}-J_{3}}{2} \)&
 0&
\( \left| \psi _{5}\right\rangle  \)=&
 \( \uparrow \uparrow \downarrow \downarrow -\downarrow \downarrow \uparrow \uparrow  \)\\
\hline
2&
 \( \frac{J_{1}+J_{4}+J_{3}}{2} \)&
 0&
\( \left| \psi _{6}\right\rangle  \)=&
 \( \uparrow \uparrow \downarrow \downarrow +\downarrow \downarrow \uparrow \uparrow +\uparrow \downarrow \uparrow \downarrow +\downarrow \uparrow \downarrow \uparrow  \)\\
&
&
&
&
\( +\uparrow \downarrow \downarrow \uparrow +\downarrow \uparrow \uparrow \downarrow  \)\\
\hline
1&
 \( \frac{-J_{1}-J_{4}+J_{3}}{2} \)&
 1&
\( \left| \psi _{7}\right\rangle  \)=&
 \( \uparrow \uparrow \uparrow \downarrow -\uparrow \uparrow \downarrow \uparrow -\uparrow \downarrow \uparrow \uparrow +\downarrow \uparrow \uparrow \uparrow  \)\\
\hline
1&
 \( \frac{-J_{1}+J_{4}-J_{3}}{2} \)&
 1&
\( \left| \psi _{8}\right\rangle  \)=&
 \( \uparrow \uparrow \uparrow \downarrow -\uparrow \uparrow \downarrow \uparrow +\uparrow \downarrow \uparrow \uparrow -\downarrow \uparrow \uparrow \uparrow  \)\\
\hline
1&
 \( \frac{J_{1}-J_{4}-J_{3}}{2} \)&
 1&
\( \left| \psi _{9}\right\rangle  \)=&
 \( \uparrow \uparrow \uparrow \downarrow +\uparrow \uparrow \downarrow \uparrow -\uparrow \downarrow \uparrow \uparrow -\downarrow \uparrow \uparrow \uparrow  \)\\
\hline
2&
 \( \frac{J_{1}+J_{4}+J_{3}}{2} \)&
1&
\( \left| \psi _{10}\right\rangle  \)=&
 \( \uparrow \uparrow \uparrow \downarrow +\uparrow \uparrow \downarrow \uparrow +\uparrow \downarrow \uparrow \uparrow +\downarrow \uparrow \uparrow \uparrow  \)\\
\hline
 2&
 \( \frac{J_{1}+J_{4}+J_{3}}{2} \)&
 2&
\( \left| \psi _{11}\right\rangle  \)=&
 \( \uparrow \uparrow \uparrow \uparrow  \) \\
\hline
\end{tabular}

\vspace{0.1cm}

\( X=\sqrt{J^{2}_{1}+J^{2}_{4}+J^{2}_{3}-J_{1}J_{4}-J_{1}J_{3}-J_{3}J_{4}} \)

Table I: The energy eigenvalues and eigenvectors of a four-spin plaquettes
with exchange interactions of strengths \( J_{1} \)(vertical), \( J_{4} \)(horizontal)
and \( J_{3} \)(diagonal). \( \frac{c_{1}}{c_{2}} \) and \( \frac{d_{1}}{d_{2}} \)
are functions of \( J_{1} \), \( J_{4} \) and \( J_{3} \). 

\vspace{0.1cm}

The eigenvalues and the eigenstates of the four-spin plaquette Hamiltonian
are shown in Table 1. For the most generalised case, the ground state
energy \( E_{jo} \) is \( \frac{-(J_{1}+J_{4}+J_{3})}{2}-X \). The
ground state wave function is of the RVB(resonating valence bond)-type.
It is a linear combination of two valance bond(VB) states with coefficients
\( c_{1} \) and \( c_{2} \) depending on the exchange interaction
strengths. Note that the other singlet-state is also of the RVB-type
with the coefficients \( d_{1} \), \( d_{2} \) depending on the
exchange interaction strengths. Some of the special cases of interest
are\\
(i) \( J_{1}=J_{4} \), \( J_{3}=0 \)

The ground state energy \( E_{j0}=-2J_{1} \) and \( \frac{c_{1}}{c_{2}}=1 \).
The other singlet state has energy zero and \( \frac{d_{1}}{d_{2}}=-1 \).\\
(ii) \( J_{1}=J_{4} \), \( J_{3}\neq 0 \)

For \( J_{3}<J_{1} \), \( E_{j0}=-2J_{1}+\frac{J_{3}}{2} \), \( \frac{c_{1}}{c_{2}}=1 \).
For \( J_{3}>J_{1} \), \( E_{j0}=-\frac{3J_{3}}{2} \), i.e., the
other RVB state becomes the ground state with \( \frac{d_{1}}{d_{2}}=-1 \).\\
(iii) \( J_{1}=J_{4}=J_{3} \)

The ground state becomes doubly degenerate. The two states have a
pair of VBs along either the horizontal or the vertical bonds, with
\( E_{j0}=-\frac{3J_{1}}{2} \).\\
(iv) \( J_{1}=J_{3}\leq \frac{J_{4}}{2} \) 

The ground state has a pair of singlets along the horizontal bonds
with \( E_{j0}=-\frac{3J_{4}}{2} \).\\
(v) \( J_{4}=J_{3}\leq \frac{J_{1}}{2} \)

The ground state has a pair of singlets along the vertical bonds with
\( E_{j0}=-\frac{3J_{1}}{2} \).\\
In the generalised as well as the special cases and for \( J_{2}\leq \frac{J^{\prime }}{4} \),
the exact ground state of the full ladder model is of the product
form. The plaquettes are in their ground state spin configurations
and the rungs are occupied by singlets.


\begin{figure}[b]
\begin{center}
\includegraphics[width=6cm]{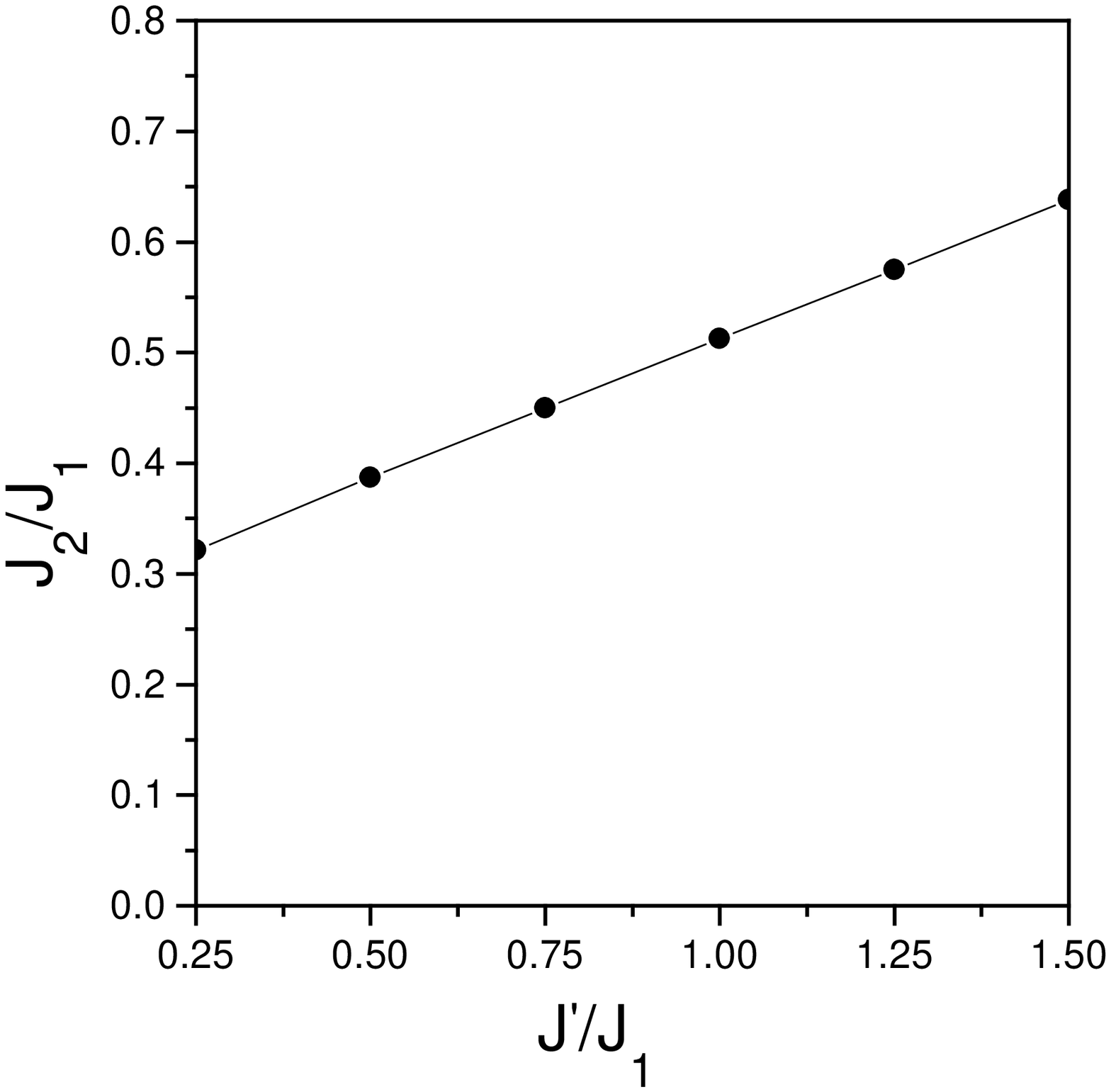}
\end{center}
\caption{Phase diagram of the ladder model (fig.1) in the parameter
space of \( \frac{J_{2}}{J_{1}} \) and \( \frac{J^{\prime }}{J_{1}} \)
for \( \frac{J_{3}}{J_{1}}=0.75 \) and \( \frac{J_{4}}{J_{1}}=0.5 \).
The parameter space below the solid line corresponds to the phase
in which the exact ground state is a product over the ground states
of the rungs and the plaquettes.}
\label{fig2}
\end{figure}


We next checked whether the exact ground state retains its product
form when \( J_{2} \) is made larger than \( \frac{J^{\prime }}{4} \).
For this, we write the total Hamiltonian \( H \) (Eq. (2)) as a sum
over six-spin sub-Hamiltonians, \( h_{i} \)'s, i.e., \( H=\sum _{i}h_{i} \).
Each sub-Hamiltonian describes a plaquette coupled to a rung. The
six-spin sub-Hamiltonian can be diagonalised exactly to obtain the
ground state energy. Again, we use the method of `divide and conquer'.
When the six-spin sub-Hamiltonians are added together to obtain the
full Hamiltonian, the \( J_{1},J_{3},J^{\prime } \) bonds are counted
twice and the \( J_{2} \) bonds only once. We identify the region
of parameter space in which the full ladder ground state has the product
form. Fig. 2 shows the phase boundaries, in the parameter space of
\( \frac{J_{2}}{J_{1}} \) and \( \frac{J^{\prime }}{J_{1}} \) for
fixed values of \( \frac{J_{4}}{J_{1}} \)(=0.5) and \( \frac{J_{3}}{J_{1}} \)(=0.75).
In the parameter regime below the phase boundary, the exact ground
state has the product form. One finds that even for \( J_{2}>\frac{J^{\prime }}{4} \),
the exact ground state has the product structure.


\begin{figure}[b]
\begin{center}
\includegraphics[width=6cm]{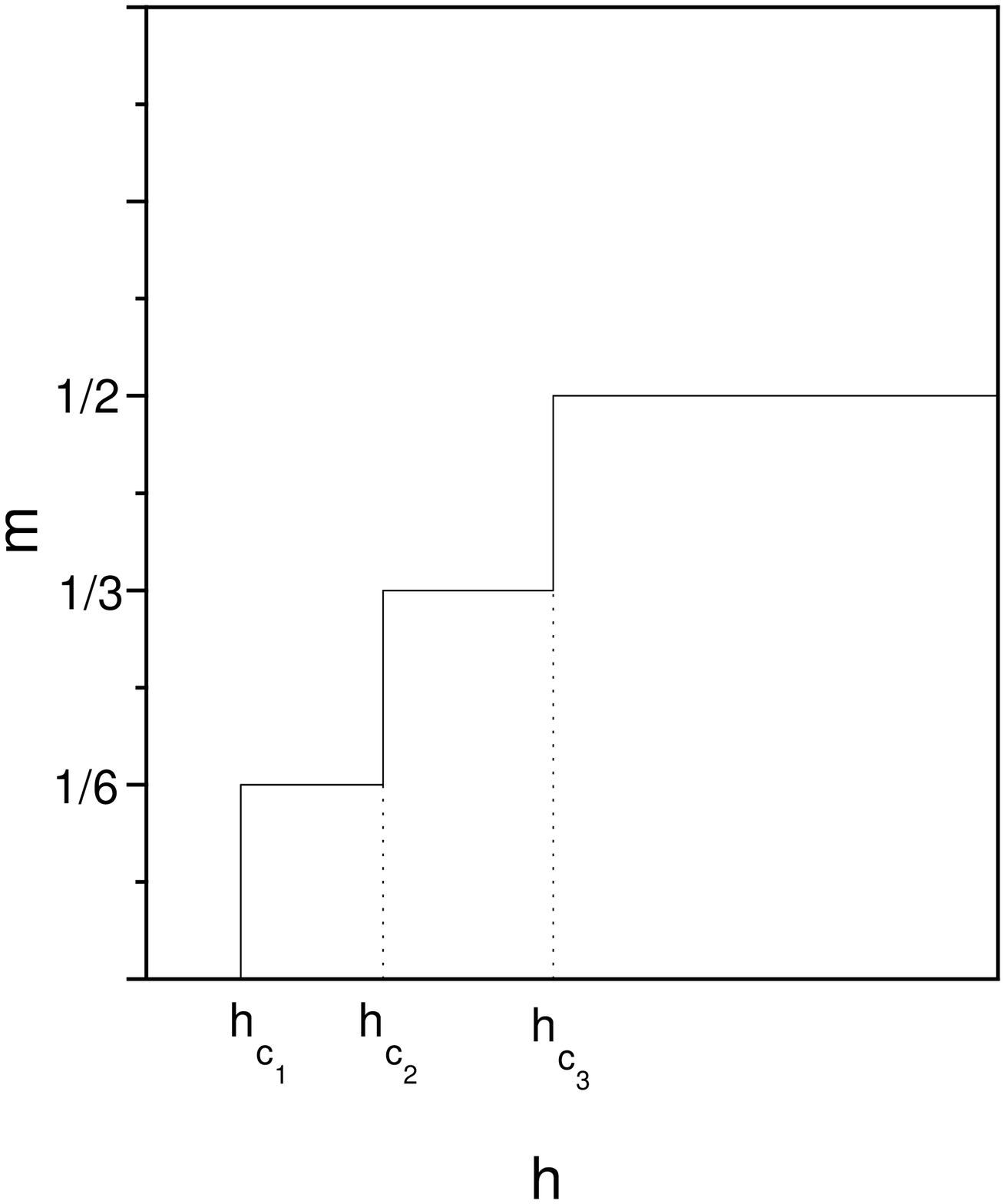}
\end{center}
\caption{Plot of magnetization/site \( m \) versus external magnetic
field \( h \) for the two-chain ladder model shown in fig. 1. The
plot is obtained in the parameter region in which the exact ground
states in different \( S_{tot}^{z} \) subspaces have the product
form. Two non-trivial magnetization plateaux occur at \( m=\frac{1}{6} \)
and \( m=\frac{1}{3} \).}
\label{fig3}
\end{figure}


\section{Magnetization plateaux}

We next include an external magnetic field term \( -h\sum ^{6N}_{i=1}S^{z}_{i} \)
in the Hamiltonian \( H \) (Eq. (2)), where 6N is the total number
of sites in the ladder. Let us first consider the case of a single
4-spin plaquette in a magnetic field. The magnetic field couples to
the z-component of the total spin of the plaquette, \( S^{z}_{tot} \),
which is a conserved quantity. The ground state energy \( E_{g}(S_{tot}^{z}) \)
at \( h=0 \) for \( S_{tot}^{z}=0,1 \) and \( 2 \) can be obtained
from Table I. When the external field \( h\neq 0 \), the ground state
in each \( S^{z}_{tot} \) subspace is \( E_{g}(S_{tot}^{z},h)=E_{g}(S_{tot}^{z},0)-hS_{tot}^{z} \).
The ground state magnetization curve can be easily obtained. The magnetization
per site \( m \) is zero from \( h=0 \) upto a critical field \( h_{c_{1}} \).
Below \( h=h_{c_{1}} \), \( S^{z}_{tot} \) of the ground state is
zero. For \( h_{c_{1}}<h<h_{c_{2}} \), \( S^{z}_{tot} \) of the
ground state is 1, so that \( m=\frac{1}{4} \) and beyond \( h=h_{c_{2}} \),
\( S^{z}_{tot} \) of the ground state is 2, i.e., the saturation
magnetization \( m=\frac{1}{2} \) is obtained. Let us first consider
the case \( J_{3}<J_{1} \). For \( J_{3}<J_{4} \), the ground state
in the \( S^{z}_{tot}=1 \) subspace is \( \left| \psi _{7}\right\rangle  \)
and the critical fields are \( h_{c_{1}}=J_{3}+X \) and \( h_{c_{2}}=J_{1}+J_{4} \).
For \( J_{3}>J_{4} \), the ground state in the \( S^{z}_{tot}=1 \)
subspace is \( \left| \psi _{8}\right\rangle  \) and \( h_{c_{1}}=J_{4}+X \)
and \( h_{c_{2}}=J_{1}+J_{3} \). At \( J_{3}=J_{4} \) these two
states are degenerate. For \( J_{3}>J_{1} \), as long as \( J_{1}<J_{4} \),
the ground state in the \( S^{z}_{tot}=1 \) subspace is \( \left| \psi _{9}\right\rangle  \)
with \( h_{c_{1}}=J_{1}+X \) and \( h_{c_{2}}=J_{4}+J_{3} \). For
\( J_{1}>J_{4} \), the ground state in the \( S^{z}_{tot}=1 \) subspace
is \( \left| \psi _{8}\right\rangle  \) with \( h_{c_{1}}=J_{4}+X \)
and \( h_{c_{2}}=J_{1}+J_{3} \). At \( J_{1}=J_{4} \), \( \left| \psi _{8}\right\rangle  \)
and \( \left| \psi _{9}\right\rangle  \) are degenerate ground states.
Hence depending on the exchange interaction strengths, we get different
values for \( h_{c_{1}} \) and \( h_{c_{2}} \). 


\begin{figure}[t]
\begin{center}
\includegraphics[width=6cm]{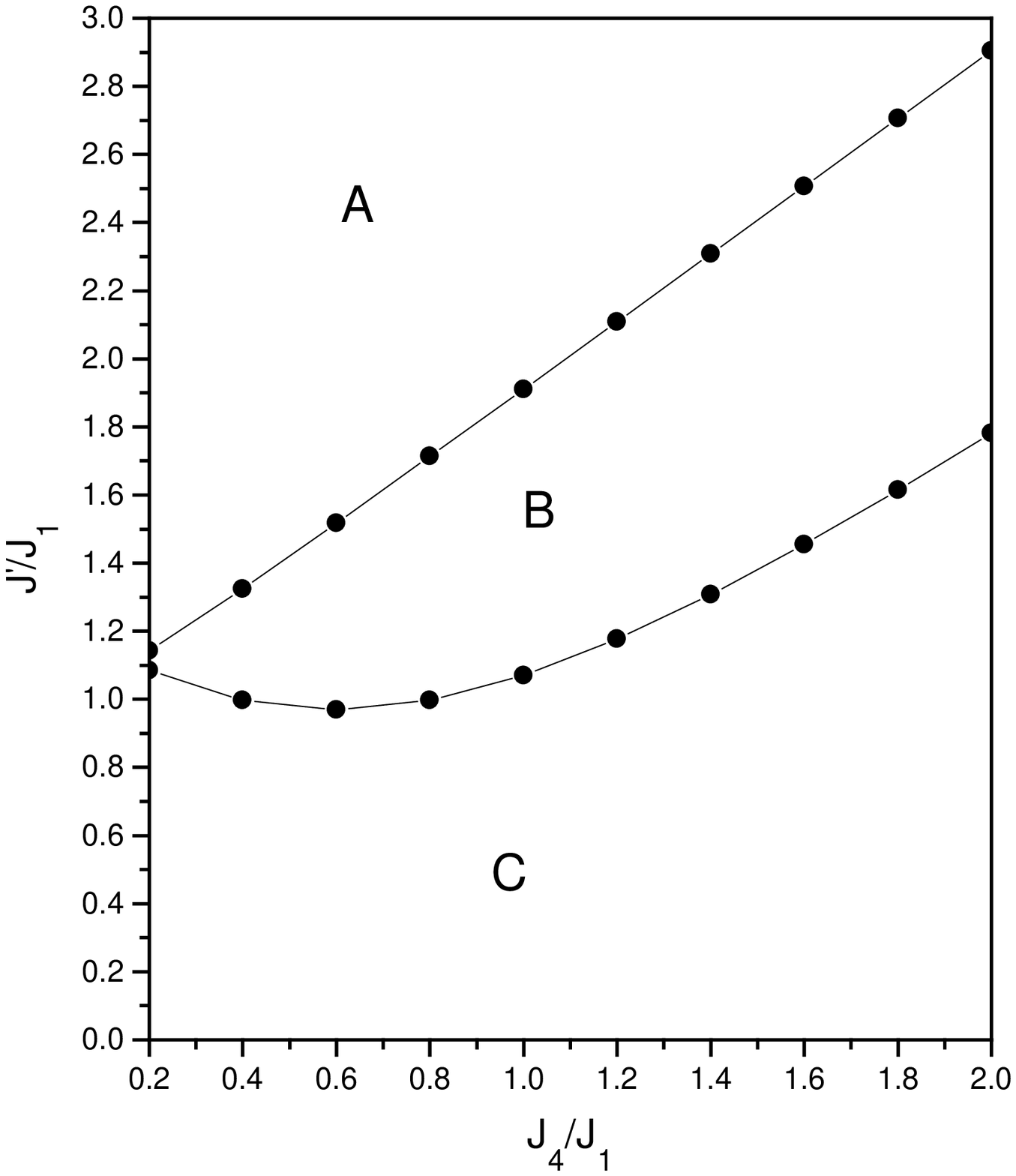}
\end{center}
\caption{Phase diagram of the ladder model (fig. 1) in a finite magnetic
field and in the parameter space of \( \frac{J^{\prime }}{J_{1}} \)
and \( \frac{J_{4}}{J_{1}} \) with \( \frac{J_{2}}{J_{1}}=0.1 \)
and \( \frac{J_{3}}{J_{1}}=0 \). The regions A, B and C are explained
in the text.}
\label{fig4}
\end{figure}



\begin{figure}[t]
\begin{center}
\includegraphics[width=6cm]{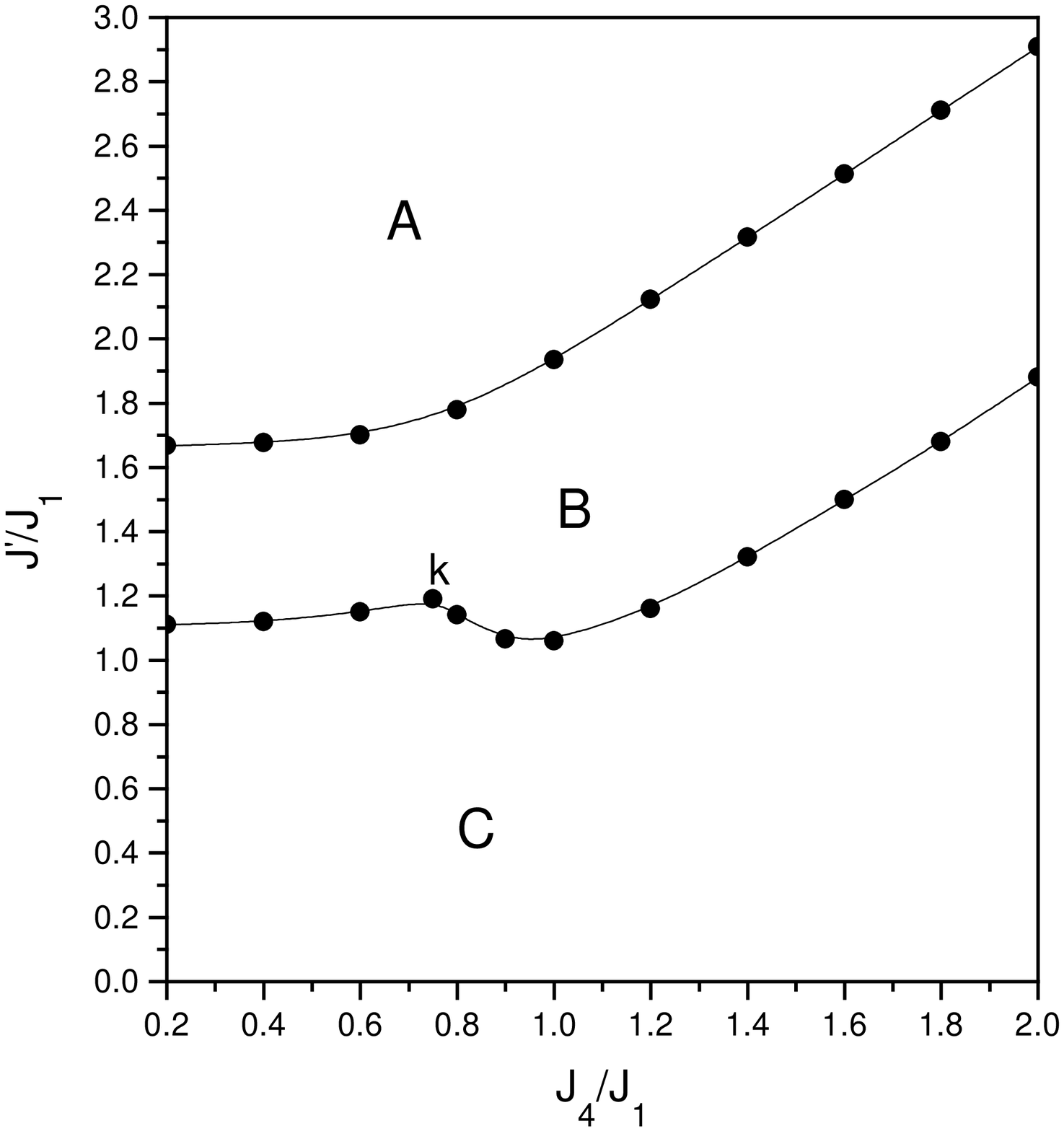}
\end{center}
\caption{Phase diagram of the ladder model (fig. 1) in a finite magnetic
field and in the parameter space of \( \frac{J^{\prime }}{J_{1}} \)
and \( \frac{J_{4}}{J_{1}} \) with \( \frac{J_{2}}{J_{1}}=0.1 \)
and \( \frac{J_{3}}{J_{1}}=0.75 \). The regions A, B and C are explained
in the text.}
\label{fig5}
\end{figure}

For the external field \( h=0 \), we have already seen that there
is an extended parameter regime in which the exact ground state of
the full ladder is a product of the ground states of the rungs and
the plaquettes. We now investigate whether the same holds true in
the presence of a finite magnetic field. Again, we use the method
of `divide and conquer' and the six-spin sub-Hamiltonian consists
of a plaquette coupled to a rung. For the full ladder, one can identify
a region (region A in Fig. 4 and Fig. 5) in parameter space in which
for \( 0<h<h_{c_{1}} \), \( m \) is zero. At \( h_{c_{1}} \), there
is a jump in the value of \( m \) to \( m=\frac{1}{6} \) and a plateau
is obtained for \( h \) upto \( h_{c_{2}} \) (Fig. 3). When \( h_{c_{1}}<h<h_{c_{2}} \),
the exact ground state has the plaquettes in their \( S^{z}=1 \)
ground states and the rungs in singlet spin configurations. Since,
the number of plaquette  is N and the total number of sites is 6N,
the magnetization/site \( m \) in the ground state is \( \frac{1}{6} \).
The quantization condition in Eq. (1) is obeyed as unit period of
the ground state contains six spins so that \( S_{u}=3 \) and the
magnetization \( m_{u} \) in the unit period is 1. At \( h_{c_{2}} \),
the second jump in \( m \) from \( \frac{1}{6} \) to \( \frac{1}{3} \)is
obtained. When \( h_{c_{2}}<h<h_{c_{3}} \), the exact ground state
has the plaquettes in their \( S^{z}=2 \) ground states and the rungs
in singlet spin configurations. In this case, \( S_{u} \) and \( m_{u} \)
in Eq. (1) are 3 and 2 respectively. At \( h=h_{c_{3}} \), there
is a jump in \( m \) from \( \frac{1}{3} \) to the saturation magnetization
\( \frac{1}{2} \). The value of \( h_{c_{3}} \), the critical field
for which the full ladder reaches saturation magnetization is \( J^{\prime }+J_{2} \).
There are other parameter regions (regions B and C in Fig. 4 and Fig.
5) in the parameter space in which the full plateau structure in the
\( m \) versus \( h \) plot, as shown in Fig. 3, is not obtained.
Fig. 4 shows the phase diagram for the full ladder in a magnetic field
in the \( \frac{J^{\prime }}{J_{1}} \) vs. \( \frac{J_{4}}{J_{1}} \)
parameter space when \( J_{3}=0 \) and \( \frac{J_{2}}{J_{1}}=0.1 \).
The region A exhibits the full plateau structure in \( m \) vs. \( h \)
as shown in Fig. 3. In region B, the jump in \( m \) from 0 to \( \frac{1}{6} \)
occurs at \( h=h_{c_{1}} \) (Fig. 3) but beyond \( h_{c_{2}} \),
the ground state is no longer of the product form. In region C, the
ground state loses its simple product structure beyond \( h=h_{c_{1}} \).
A similar phase diagram is shown in Fig. 5, for \( \frac{J_{3}}{J_{1}}=0.75 \),
and with all other parameters the same as in the case of Fig. 4. We
observe a kink at the point k, where \( J_{4}=J_{3} \) and a transition
from the phase B to phase C occurs. At this point, the ground state
in \( S^{z}_{tot}=1 \) subspace changes from \( \left| \psi _{8}\right\rangle  \)
to \( \left| \psi _{9}\right\rangle  \). The kink is indicative of
phase reentrance. When \( \frac{J^{\prime }}{J_{1}} \) is in the
range 1.11<\( \frac{J^{\prime }}{J_{1}} \)<1.19, one gets the phases
B-C-B-C as \( \frac{J_{4}}{J_{1}} \) is varied (\( \frac{J_{4}}{J_{1}}\geq 0.2 \)).
Similarly, for 1.06<\( \frac{J^{\prime }}{J_{1}}\leq  \)1.11, the
phases C-B-C are obtained.

\section{Concluding remarks}

In this paper, we have introduced a generalised ladder model with
modulated exchange interactions. The model consists of four-spin plaquettes
coupled to two-spin rungs. In a wide parameter regime, the exact ground
state of the model is a product over the ground states of the individual
plaquettes and rungs. In the presence of an external magnetic field,
magnetization plateaux are obtained when magnetization/site m is plotted
as a function of the external field h. In an extended parameter regime,
the exact ground states in the different magnetization subspaces are
of the product form. The generalised model includes several other
ladder models as special cases. For \( J_{2}=J_{3}=J_{4}=J \) and
\( J^{\prime }=J_{1} \), the model reduces to the frustrated ladder
model introduced by Bose and Gayen{[}19{]}. As Xian{[}20{]} has shown,
for \( \frac{J^{\prime }}{J}>(\frac{J^{\prime }}{J})_{c}\simeq 1.40148, \)
the exact ground state consists of singlets along the rungs of the
ladder. At \( \frac{J^{\prime }}{J}=(\frac{J^{\prime }}{J})_{c} \),
there is a transition from the rung dimer state to the Haldane phase
of the S=1 chain. Similar arguments show that for \( J_{4}=J_{3} \)
in our generalised ladder model, the ground state in a certain parameter
regime is that of a spin one chain with modulated exchange interactions.
The sequence of exchange couplings along the chain has the structure
\( J_{4}-J_{2}-J_{2}-J_{4}-J_{2}-J_{2}\cdots \cdots  \). Other examples
of ladder models which are special cases of the generalised model,
have been given in Section 2. Further studies are needed to obtain
the phase diagram of the generalised model in the full parameter space.

\subsection*{Acknowledgment}

E. Chattopadhyay is supported by the Council of Scientific and Industrial
Research, India under sanction No. 9/15(186)/97-EMR-I.E.


\begin{thebibliography}{10}
\bibitem{key-1}E. Dagotto and T. M. Rice, Science 271, 618 (1996) 
\bibitem{2}E. Dagotto, Rep. Prog. Phys. 62, 1525 (1999) 
\bibitem{key-9}D. C. Cabra, A. Honecker and P. Pujol, Phys. Rev. Lett. 79, 5126 (1997);
Phys. Rev.B 58, 6241 (1998) 
\bibitem{6}K. Tandon, S. Lal, S. K. Pati, S. Ramasesha and D. Sen, Phys. Rev.
B 59, 396 (1999) 
\bibitem{7}F. Mila, Eur. Phys. J. B 6, 201 (1998) 
\bibitem{8}D. C. Cabra, M. D. Grynberg, Phys. Rev. Lett. 82, 1768 (1999) 
\bibitem{9}D. C. Cabra, M. D. Grynberg, A. Honecker and P. Pujol, cond-mat/0010376
and references therein. 
\bibitem{10}T. Hakobyan, J. H. Hetherington and M. Roger, Phys. Rev. B 63, 144433
(2001)
\bibitem{11}W. Brenig, K. W. Becker and P. Lemmens, cond-mat/0101460; W. Brenig
and K. W. Becker, cond-mat/0105096 
\bibitem{12}M. Johnsson, K. W. T\"{o}rnross, F. Mila and P. Millet, Chem. Mater.
12, 2853 (2000) 
\bibitem{key-5}A. Honeker, F. Mila, M. Troyer, Eur. Phys. J. B15 227 (2000) 
\bibitem{key-6}E. M\"{u}ller-Hartmann, R.R.P.Singh, C. Knetter, G.S. Uhrig, Phys.
Rev. Lett. 84 1808 (2000) 
\bibitem{key-7}A. Koga, K.Okunishi, N. Kawakami, Phys. Rev B 62 5558 (2000) 
\bibitem{key-8}J. Schulenburg and J. Richter, Phys. Rev. B 65, 054420 (2002)
\bibitem{key-10}O. Kahn, Molecular Magnetism (VCH Publisher, New York 1993) 
\bibitem{key-11}M. Oshikawa, M. Yamanaka and I. Affleck, Phys. Rev. Lett. 78, 1984
(1997)
\bibitem{key-14}E. Chattopadhyay and I. Bose, Phys. Rev. B. 65, 134425, 2002
\bibitem{key-12}P. W. Anderson, Phys. Rev. 83, 1260 (1951) 
\bibitem{14}I. Bose and S. Gayen, Phys. Rev. B 48, 10653 (1993) 
\bibitem{key-13}Y. Xian, Phys. Rev. B 52, 12485 (1995) 
\end{thebibliography}
\end{document}